\documentclass[aps,prl,superscriptaddress,amsmath,amssymb,floatfix,twocolumn]{revtex4}
\usepackage{graphicx}  
\usepackage{subfigure}
 \usepackage{epsfig} 
 \usepackage{bm}
\begin{document}
%%%%%%%%%%%%%%%%%% title page information %%%%%%%%%%%%%%%%%%
\title{Mode Bifurcation and Fold Points  of Complex Dispersion Curves for the Metamaterial Goubau Line}
\author{P. L. Overfelt}
\author{Klaus Halterman}
\email{klaus.halterman@navy.mil}
\author{Simin Feng}
\author{D. R. Bowling}
\affiliation{Research and Intelligence Department, Physics Division, Naval Air Warfare Center, China Lake, California 93555, USA}
\date{\today}

%%%%%%%%%%%%%%%%%%% abstract and OCIS codes %%%%%%%%%%%%%%%%

\begin{abstract}
In this paper the complex dispersion curves of the four lowest-order transverse magnetic modes
of a dielectric Goubau line ($\epsilon>0, \mu>0$) are compared with those of a dispersive metamaterial
Goubau line. The vastly different dispersion curve structure for the metamaterial Goubau line is characterized
by unusual features such as mode bifurcation, complex fold points,
both proper and improper complex modes, and merging of complex and real modes.
\end{abstract}
\maketitle
The %structure called the 
Goubau line (G-Line) has been known and studied since Sommerfeld and later Goubau 
considered applications using non-radiating surface waves 
on transmission lines \cite{sommer,gball}. While Sommerfeld analyzed a long cylindrical metallic 
wire as the transmission line of interest, Goubau realized that adding a 
dielectric outer sheath to the wire reduced the radial extent of the electromagnetic (EM) field and thus 
the dimensions of the associated excitation device. 
Interestingly, while the Sommerfeld wave can exist only on a conductor of finite conductivity, 
the Goubau wave can exist even when the inner conductor is assumed 
to have perfect conductivity. The G-Line has been investigated since the first part of the 
twentieth century, and its guided modes are well known \cite{sommer,gball,strat,wald,semen}. 
As with all open waveguide structures, the G-Line spectrum consists of a finite discrete set 
of guided modes with purely real longitudinal propagation 
constants and an infinite continuum of radiation modes. Also present on open lossless structures 
are leaky waves \cite{marc,marc2} characterized by discrete complex 
longitudinal propagation constant solutions to the dispersion equation, but which are improper 
solutions of Maxwell's equations in that these solutions 
decay longitudinally but do not obey the transverse radiation condition and thus may only be used in 
restricted regions of space. Improper waves are not 
considered part of an open waveguide spectrum and are often referred to as ``nonmodal" or 
``nonspectral" \cite{marc}.   Despite this fact, leaky waves have found 
great usefulness in certain applications, particularly those related to leaky wave antennas \cite{gold}.  
Some authors have referred to the complex solutions of 
the circular dielectric rod and the standard G-Line dispersion equations as  {\it leaky modes} and 
have considered them on a more or less equal footing 
with the guided modes \cite{kim,kim2}. The leaky waves of even the standard G-Line are still not well 
characterized (only the transverse magnetic (TM) solutions have been considered 
in detail \cite{kim}) but on that structure, all complex leaky wave solutions of the characteristic equation 
have EM  field components that diverge as 
the radial coordinate increases to infinity and are thus improper. 

\begin{figure}
\centering
\includegraphics[width=.3\textwidth]{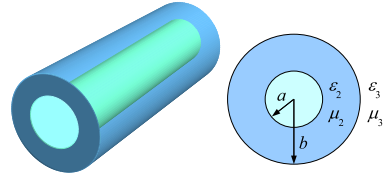}
\caption{Schematic of the Goubau line geometry.
The metal core of radius $a$ is a perfect conductor.
}
\label{f1} % caption for the whole figure
\end{figure}
In this Letter the G-Line geometry (see Fig.~\ref{f1}) is used with a negative index 
of refraction dispersive metamaterial 
(NIM) \cite{elef,shad} replacing the usual dielectric layer. 
Under these circumstances the metamaterial G-Line spectrum consists of 
guided modes, radiation modes, improper complex waves, and {\it proper} 
complex modes. 
%The latter are characterized by electromagnetic 
%field components that obey the radiation condition. 
In the following we consider only the
symmetric transverse magnetic (${\rm TM}_{0n}$) 
%and 
%transverse electric (${\rm TE}_{0n}$) 
solutions of the metamaterial G-Line (MM G-line). 
The transverse electric (${\rm TE}_{0n}$) and
hybrid modes of the MM G-line will be considered 
subsequently. 
%Throughout the following, the 
%$\epsilon$ positive, $\mu$ positive dielectric Goubau line will be 
%referred to as the G-line or standard G-line.

Using the geometry in Fig.~\ref{f1}, the characteristic 
equation for the G-Line with perfectly conducting inner conductor and assuming a
$\theta,z,t$ dependence of $e^{-i(m\theta+\gamma z- \omega t)}$, can be written in the general form,
\begin{align}
\overline{\gamma}^2 m^2 \Bigl(\frac{1}{s^2}-\frac{1}{w^2}\Bigr)^2 
{\cal Z} {\cal Y} + {\cal S} {\cal T} = 0,
\label{disp}
\end{align}
where $\overline{\gamma} = \overline{\beta} + i \overline{\alpha}$ is the normalized
complex longitudinal propagation constant.
The longitudinal phase, $\overline{\beta}$, and attenuation $\overline{\alpha}$ are both normalized by $k_0 \equiv \omega/c$, the free space wavenumber.
Considering  the $m = 0$ case, Eq.~(\ref{disp}) reduces to ${\cal T} = 0$  
for TM modes, %(or ${\cal S} = 0$ for TE modes), 
where, 
\begin{align}
{\cal T} = \frac{\epsilon_2 {\cal W}}{w} - \epsilon_3 {\cal Z} g =0, 
\label{calT}
\end{align}
%For ${\rm TE}_{0n}$ modes,
%\begin{align}
${\cal S} = {\mu_2 {\cal X}}/{w} - \mu_3 {\cal Y} g$, % =0,
%\label{calS}
%\end{align}
and $g\equiv -H_1^{(2)}(s)/(s H_0^{(2)}(s))$.
We also define,
${\cal Z} = J_0(w)Y_0(v)-J_0(v)Y_0(w)$,
${\cal Y} = -J_0(w) Y_1(v) + J_1(v)Y_0(w)$,
${\cal W} = -J_1(w) Y_0(v) + J_0(v) Y_1(w)$,
and ${\cal X} = J_1(w)Y_1(v)-J_1(v)Y_1(w)$,
with the $J$'s, $Y$'s and $H$'s the usual
Bessel functions of the first and second kind, and the Hankel
functions of the second kind, respectively.
The dimensionless wavenumbers are,
$v=k_2 a$, $w=k_2 b$, and $s=k_3 b$.
The transverse propagation wave numbers are given by
$k_j = k_0 \sqrt{\epsilon_j \mu_j - \overline{\gamma}^2}$, for $j=2,3$.
For these ${\rm TM}_{0n}$ modes,
the EM fields in the $j$th region are generally  written as, ${\bm H}_j = H_{j,\theta} \hat{\bm \theta}e^{-i (\gamma z- \omega t)}$,
and ${\bm E}_j = (E_{j,\rho} \hat{\bm \rho} + E_{j,z} \hat{\bm z})e^{-i (\gamma z- \omega t)}$,
for propagation in the positive $z$ direction.
Maxwell's equations and the cylindrical symmetry result in the following transverse fields, 
$H_{j,\theta} = -i \epsilon_j k_0/\kappa_j^2 \partial E_{j,z}/\partial \rho$, and 
$E_{j,\rho}=-i \gamma /\kappa_j^2 \partial E_{j,z}/\partial \rho$.
%, and a number of modified Goubau line 
%configurations have been considered also \cite{rao}. 
The forms of (\ref{disp}) - (\ref{calT}) above are most suitable for emphasizing material 
parameter changes in both $\epsilon$ and $\mu$ of Regions 2 and 3,
although they have been derived previously in alternate forms \cite{semen}.

\begin{figure}
\centering
\includegraphics[width=.27\textwidth]{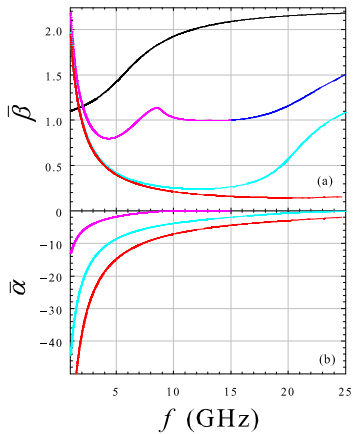}
\caption{Dispersion for standard G-line
}
\label{f2} % caption for the whole figure
\end{figure}
We solve
Eq.~(\ref{calT}) numerically for its complex zeros as a function of frequency by
employing a recently developed complex root finding algorithm.  
%Put in a description of the algorithm. 
Fig.~\ref{f2} shows the four lowest-order ${\rm TM}_{0n}$ ($n=0,1,2,3$) modes as functions of frequency 
(in GHz) for the standard G-Line. Part (a) shows the normalized longitudinal phase vs. frequency while Part (b) shows normalized 
longitudinal attenuation vs. frequency. In Fig.~\ref{f2} the parameters $\epsilon_2=5$, $\mu_2=1$, $a= 5$ mm, $b=10$ mm were used to compare our results 
with those in Ref.~\onlinecite{kim}. In particular, note the lowest-order ${\rm TM}_{00}$ mode (black curve)
that has no cutoff frequency and $\alpha=0$. Except for this 
mode, our results are consistent with Ref.~\onlinecite{kim}. We have assumed that the outside region is 
air and thus $\epsilon_3=\mu_3=1$. At a given frequency, when the imaginary part of a root is zero, that root represents a guided or bound 
solution. When the imaginary part is nonzero, that root represents a complex leaky wave solution of (\ref{calT}). Upon calculating the EM 
field components associated with certain leaky wave solutions in Fig.~\ref{f2}, all discrete complex solutions of the standard G-line 
(with Region 2 index of refraction greater than one) are improper \cite{marc,kim}. 
Because the outside region is air, cutoff occurs when $\overline{\gamma} = 1$ 
or $s = 0$. When  $1<\overline{\beta} <\sqrt{5}$,  
and  $\overline{\alpha} = 0$, the well-known guided modes result. 
When  $\overline{\beta} < 1$, $\overline{\alpha}\neq 0$, the 
waves are leaky. Also note that an infinite number 
of higher order leaky waves can occur at 
lower frequencies 
%(we have shown only a small number). 
These leaky
waves have high frequency cutoffs 
but no low frequency cutoffs. Generally these waves have 
$|\overline{\alpha}| \gg 1$ and thus
are not so far found to be particularly useful. 
In Fig.~\ref{f2}, the real and imaginary parts of the longitudinal propagation 
constant are correlated using color. Thus the lowest-order mode (black) has zero 
attenuation and is purely guided [Fig.~\ref{f2}(a)], the second lowest-order mode (magenta) 
is leaky at the lower frequencies with its attenuation magnitude increasing as the 
frequency decreases (see Fig.~\ref{f2}(a) and (b)). For this mode as the frequency increases up 
to cutoff at about 14.9 GHz, the associated attenuation decreases in magnitude until this 
mode becomes a guided mode once cutoff is passed (magenta $\rightarrow$ blue). The turquoise blue 
and red curves show the next higher-order modes which for much of the frequency range
are leaky with $|\alpha| \gg 1$.

If the dielectric layer of Region 2 is replaced by a negative index of refraction MM, the TM mode structure is vastly different. 
The MM is characterized by the following dispersion formulas, 
$\mu_2(f) = \mu_l + F f^2/(f_0^2-f^2-i \Gamma f)$, and, $\epsilon_2(f) = \epsilon_h + f_p^2/(f_r^2-f^2-i\Gamma f)$.
We take the filling factor, $F$, to be $0.25$, and
the damping factor, $\Gamma$, to be  zero. The low and high frequency limits are
$\mu_l = -1$ and $\epsilon_h = -5$, respectively.
The resonance frequencies are, $f_0 = 0.75$ GHz,
$f_r = 0.5$ GHz, and
$f_p = 0.25$ GHz.

\begin{figure}
\centering
\includegraphics[width=.4\textwidth]{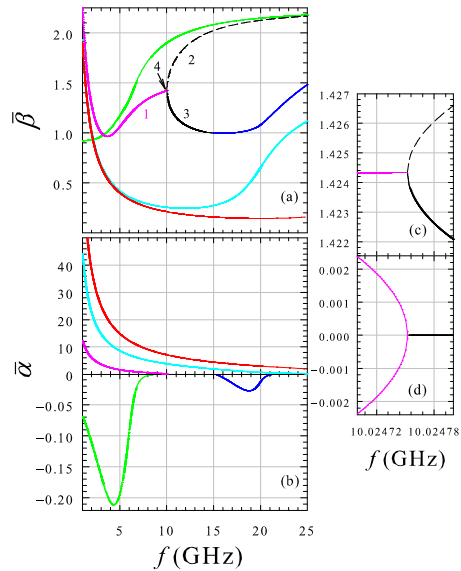}
\caption{Dispersion for the MM G-line.
The curves  in (c) and (d) give a magnified view of $\overline{\beta}$ 
and $\overline{\alpha}$, respectively, near the bifurcation point (Point 4). 
}
\label{f3} % caption for the whole figure
\end{figure}
Figure \ref{f3} contains the four lowest-order TM modes assuming a metamaterial G-Line with both $\epsilon_2$
and $\mu_2$ negative over the frequency range of interest. Fig.~\ref{f3}(a) shows the normalized 
longitudinal phase vs. frequency, 
part (b) shows the normalized longitudinal attenuation vs. frequency, where
$|\overline{\alpha}|$ is small, and  also  shows the normalized longitudinal attenuation on a larger scale.
In Fig.~\ref{f3}, $\epsilon_2\approx -5$ and $\mu_2 \approx -1$. 
The radii, $a$ and $b$, are the same as for the standard G-line in Fig.~\ref{f2}.
Referring to the  various mode ``paths" 
or ``branches" in Fig.~\ref{f3}, note that 
%the lowest-order TM mode (along path 5) now has a leaky part 
the green curve now has a leaky part 
(below about 3 GHz) and the 
portion of the green curve with
%path from 8 to 5 above  
$\overline{\beta}>1$ 
now has  $\overline{\alpha}\neq 0$. 
As the frequency decreases, the attenuation peaks in magnitude between 4 and 5 GHz (green curve) 
and then starts to decrease.
Thus the green path is 
a wave with a complex $\overline{\gamma}$. 
This path is also improper (until it merges with the black curves
at much higher frequency). 
Thus the fundamental guided mode of the standard G-line is replaced by a complex mode 
that is improper. % (at least at the lower frequencies). 
The mode represented by the magenta path
%path 7 to 8 and on up past point 1 to point 4 
is also complex 
with $\overline{\alpha}$ nonzero until point 4 is reached. However this path 
is proper. At point 4 ($f \approx 10.025$ GHz), there is a mode bifurcation [see Fig.~\ref{f3}(c) and (d)],
as has been observed in catastrophe and bifurcation theories. Catastrophe theory 
attempts to study how the qualitative nature of solutions of equations depends on the 
parameters appearing in those equations \cite{hanson,poston}. These techniques are invaluable in 
analyzing the qualitative changes in system dynamical characteristics by small perturbations. 
From this standpoint, the dispersion relationship, ${\cal T} = 0$, is cast as a nonlinear equation 
in $\overline{\beta}$, $\overline{\alpha}$, and $f$ with parameters $\epsilon_1$,
$\epsilon_2$, $\mu_1$, $\mu_2$, $a$, and $b$. Using catastrophe 
theory,  areas of unusual qualitative interest such as the fold point at Point 4 can be determined.  
Using numerical techniques alone to solve the dispersion relation over a frequency range is 
difficult due to its ``dense" set of zeros at each frequency.
Point 4 is a 
complex fold point (a point where, because no material losses in the line 
have been assumed as yet,  $\overline{\gamma}$,  $-\overline{\gamma}$,  $\overline{\gamma}^*$, and $-\overline{\gamma}^*$ 
all meet at the same frequency) \cite{yak,hanson,poch,poston}. As the frequency increases past 
the fold point, $\overline{\alpha}\rightarrow 0$ along both upper and lower black paths. The black paths near points 2 and 3 are 
therefore both guided modes immediately after bifurcation. 

Thus, in Fig.~\ref{f3} there are an improper complex set of roots, a proper complex set of roots, and a 
mode bifurcation at a fold point in (a) near $f\approx 10$ GHz past which 
there are two guided modes (i.e., $\overline{\alpha}= 0$ on both black paths after 10 GHz). 
By considering the two guided paths (2 and 3) where  $\overline{\alpha}= 0$, 
we find 
that the black upper path always has a slope opposite in sign to the 
lower path until the first cutoff frequency, 
$f\approx  14.9$ GHz, and thus is backward traveling in this frequency region. At cutoff the lower 
path near point 3 has zero slope and after cutoff, 
the slope becomes positive. 
Also after cutoff, this path becomes complex with $|\overline{\alpha}| \ll 1$, 
as shown by the blue curve in Fig.~\ref{f3}(b). The red 
and turquoise blue curves are quite similar to those in Fig.~\ref{f2} over the 
given frequency range but they, too, exhibit the bifurcation and fold point phenomena at higher frequencies
None of the above situations occur for the standard G-line.

\begin{figure}
\centering
\includegraphics[width=.35\textwidth]{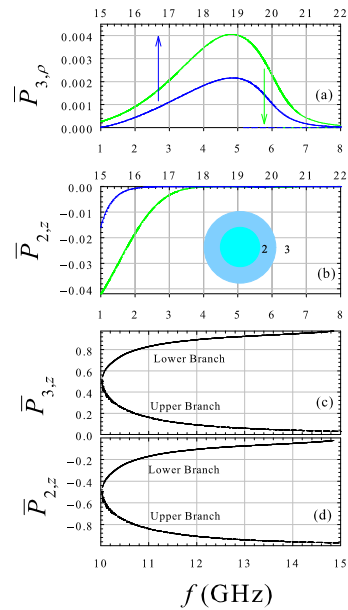}
\caption{Normalized power (see text) as a function of frequency
for a few of the dispersion curves in Fig.~\ref{f3}, and using
the same color scheme. In (a), the radial power flux increases, consistent
with the attenuation growth for the mode. 
In (b), the longitudinal power flux in the MM region declines rapidly with frequency. 
The modes in (c) and (d)
are strictly guided, with the EM energy flowing in the MM mainly for those modes
in the upper branch of the dispersion curve of Fig.~\ref{f3}.
}
\label{f4} % caption for the whole figure
\end{figure}

Certain effects similar to those in Fig.~\ref{f3} have been 
seen in both grounded metamaterial slab geometries \cite{shu,lovat} 
and in  complicated grounded 
dielectric-loaded, open geometries  \cite{yak,bac,lamp,maj,tamir}.  
In the case of the MM G-line, we have a transition from a complex proper mode
(magneta curve in Fig.~\ref{f3}(a)) 
to a pair of guided modes at higher frequencies.
We see in both the normalized phase and attenuation 
the characteristic intersection of a parabola and a straight line [Fig.~\ref{f3}(c) and (d)]. 
The phase has a straight line to parabola transition 
characteristic of a fold point in catastrophe theory \cite{yak,hanson,poch,poston}. 
Note that the attenuation for the MM G-line in Fig.~\ref{f3}(b) has changed sign for the red, 
magenta, and turquoise blue curves when compared to the standard G-line.
while the attenuation shows 
a parabola to straight line transition.

The guided and leaky modes of the G-line and MM G-line can be 
described as slow modes and fast modes, respectively. The guided modes are slow in that they 
occur in the spectral region $\sqrt{ \epsilon_3 \mu_3} < \overline{\beta} < \sqrt{\epsilon_2 \mu_2}$. 
Since $\overline{\beta}>1$, 
$c \beta/\omega >1$ but $\omega/(c \beta) <1$, thus their phase velocity is less than the speed of light. 
In contrast the leaky waves are fast waves. Note that with reference to both Fig.~\ref{f2} and Fig.~\ref{f3} there 
is a point at the low frequency end of each graph where (although we cut the graphs off at 
$\sqrt {\epsilon_2 \mu_2}$) 
there are complex solutions of the dispersion relation that have $\overline{\beta} 
> \sqrt{\epsilon_2 \mu_2}$, and $\overline{\alpha}\neq0$. 
For the standard G-line this region is a ``forbidden" region.  
For the MM G-line of Fig.~\ref{f3}, 
$\overline{\beta} > \sqrt{\epsilon_2 \mu_2}$ 
is no longer considered forbidden, and ``superslow modes" 
can arise \cite{shad}.  

The total power through the various regions can be described by
the normalized power in the radial and longitudinal directions,
$\overline{P}_{j,\sigma} = P_{j,\sigma}/P_{\rm tot}$, 
for $\sigma=\rho,z$, and $j=2$ or $3$. 
The total power, $P_{\rm tot}$ is the sum of
power flux, defined as,
$P_{\rm tot} \equiv \sum_{j,\sigma} |P_{j,\sigma}|$.
The power flux
through the surface of a  cylinder of radius $R$  and length $L$ encompassing a segment of the structure  is determined 
from the spatially integrated time-averaged Poynting vector ${\bm S}_j$, with components, 
${S}_{j,z}(\rho,z)= (c/8\pi)e^{2\alpha z}  {\rm Re} [E_{j,\rho}  H_{j,\theta}^*]$,
or ${S}_{3,\rho}(\rho,z)= -(c/8\pi) e^{2\alpha z} {\rm Re} [E_{3,z}  H_{3,\theta}^*]$.
Thus, the longitudinal power transmitted through the two circular regions  is, 
$P_{j, z}|_{z=L} =  2\pi e^{2\alpha L}\int_0^R  \rho d \rho S_{j, z}(\rho,0)$, and
$P_{j, z}|_{z=0} =  - 2\pi \int_0^R  \rho d \rho S_{j, z}(\rho,0)$.
For the surface normal to $\hat{\bm \rho}$, 
integration gives,
$P_{3, \rho} =  2\pi R \, e^{\alpha L}(1/\alpha)\sinh(\alpha L)  S_{3, \rho}(R,0)$.
We take $R/b=10$.
In Fig.~\ref{f4}, we illustrate the normalized power as a function of
frequency for several of the modes shown in Fig.~\ref{f3}.
In panel (a), 
power flows radially,
with maxima at frequencies correlated to
when $|\alpha|$ peaks for that mode (Fig.~\ref{f3}(b)).
Although the power flow is predominately in the air region,
is is evident that the counter-directed longitudinal power flux in the MM
region
declines rapidly towards zero with increased frequency [panel (b)].
This indicates that power emission from the
waveguide takes place at a slight angle, $\vartheta$ from the $z$ axis
over a relatively small bandwidth.
As for the proper mode spectra, $|\overline{P}_{j,z}|=1/2$ for both regions, but
counter-directed. 
At the bifurcation point ($f\approx 10$ GHz),
the leaky wave transforms into a strictly guided mode at higher frequencies, 
where the two counter propagating regions are clearly identified by their sign [Fig.~\ref{f4}(c) and (d)].
The lower branch (see Fig.~\ref{f3}) in the guided mode dispersion clearly maps to power
flow that is predominately in the air region.
This is in contrast to modes occupying the upper branch in Fig.~\ref{f3}, where
again in Fig.~\ref{f4}(c) and (d) we see the power distribution shift from both regions equally to
reside completely in the MM at higher frequencies.
This also verifies that the MM G-line has the remarkable
property that certain guided modes can propagate below the first waveguide cutoff and can
be of the backward as well as of the forward wave type \cite{hrab,hrab2,lub}.

In conclusion, dispersion curve behavior for a dielectric versus 
MM G-Line has been shown to be vastly different. The MM
G-line dispersion is characterized by unusual mode bifurcations, complex
fold points a la catastrophe theory, and both proper and improper complex modes.

The authors acknowledge G. A. Lindsay,
Z. Sechrist, and G. Ostrom
for valuable discussions and the 
support from ONR, 
as well as NAVAIR's ILIR program from ONR.
This work is also supported in part by a grant of HPC resources 
as part of the DOD HPCMP.

\end{document}